\documentclass[twocolumn, tighten, times, twocolappendix]{aastex7}

\begin{document}

\title{Searching for the Third Wheel: High-Contrast Imaging Constraints on Tertiaries to\\ Black Hole and Neutron Star Binaries}

\author[orcid=0000-0002-1386-0603,gname=Pranav,sname=Nagarajan]{Pranav Nagarajan}
\affiliation{Department of Astronomy, California Institute of Technology, 1200 E. California Blvd., Pasadena, CA 91125, USA}
\email[show]{pnagaraj@caltech.edu}  

\author[orcid=0000-0002-6871-1752,gname=Kareem,sname=El-Badry]{Kareem El-Badry}
\affiliation{Department of Astronomy, California Institute of Technology, 1200 E. California Blvd., Pasadena, CA 91125, USA}
\email{kelbadry@caltech.edu}  

\author[orcid=0000-0002-1838-4757,gname=Aniket,sname=Sanghi]{Aniket Sanghi}
\altaffiliation{NSF Graduate Fellow}
\affiliation{Department of Astronomy, California Institute of Technology, 1200 E. California Blvd., Pasadena, CA 91125, USA}
\email{asanghi@caltech.edu}  

\begin{abstract}
Black holes (BHs) and neutron stars (NSs) with low-mass stellar companions challenge traditional isolated binary evolution models, motivating hierarchical triple evolution as a promising alternative. To search for tertiaries, we perform deep, adaptive optics-assisted, near-infrared imaging of five quiescent BH low-mass X-ray binaries (LMXBs), Gaia BH1, and twelve Gaia NSs. We detect several faint stars previously unresolved in survey imaging, but none are close enough to robustly rule out a chance alignment. To achieve high contrast sensitivity at close separations, we use the reference star differential imaging strategy with the Karhunen-Lo\'eve Image Processing algorithm to model and subtract the point-spread function of each target. We identify tertiary candidates in the speckle-dominated regime, but injection-recovery tests suggest most 5$\sigma$ detections are likely artifacts. We derive $5\sigma$ contrast curves and convert these to limits on the mass of main sequence (MS) tertiaries and the effective temperature of white dwarf (WD) tertiaries consistent with a non-detection. We rule out plausible MS tertiaries and young, hot WD tertiaries at projected separations $\gtrsim 500$~au for the Gaia compact object binaries and $\gtrsim 2000$~au for the more distant BH LMXBs. While the recent discovery of a $1.2\,M_{\odot}$ tertiary to V404~Cygni supports triple formation scenarios for BH LMXBs, our results suggest such companions are relatively rare. Our observations remain consistent with intermediate-mass tertiaries that have since evolved into cool WDs, detectable with deeper JWST imaging. Follow-up observations are required to measure proper motions and confirm or rule out physical association of tertiary candidates.

\end{abstract}



\section{Introduction}
\label{sec:intro}

A significant fraction of black hole (BH) and neutron star (NS) progenitors are born in triples and higher order multiples \citep{moe_mind_2017, offner_multiple_2023}, suggesting that BHs and NSs may still reside in hierarchical triples today. Indeed, triple evolution can solve some of the challenges facing isolated binary evolution formation models for BHs and NSs orbited by low-mass stars, both for close, accreting systems and their wide, detached counterparts \citep[e.g.,][]{naoz_2016, generozov_perets_2024, shariat_cygni_2025, naoz_links_2025}.

The close, accreting systems are low-mass X-ray binaries (LMXBs), which host BHs or NSs accreting from $\lesssim3\,M_{\odot}$ stellar companions \citep[e.g.,][]{remillard_x-ray_2006}, with 21 Galactic BH LMXBs dynamically confirmed to date \citep[e.g.,][]{corral-santana_blackcat_2016}. While the majority of all known BHs and NSs with luminous companions are in LMXBs \citep[e.g.,][]{bahramian_low-mass_2023}, theoretical models predict that these systems are likely a rare outcome of binary evolution \citep[e.g.,][]{portegies_zwart_formation_1997}, and empirical arguments suggest that they are significantly outnumbered by wide binaries hosting non-accreting, ``dormant'' BHs and NSs \citep[e.g.,][]{el-badry_gaias_2024}. The \textit{Gaia} mission \citep{gaia_collaboration_gaia_2016, gaia_collaboration_gaia_2023} has recently uncovered this hidden population, with \textit{Gaia} astrometry already enabling the discovery and characterization of three dormant BHs (``Gaia BHs'') \citep{el-badry_sun-like_2023, chakrabarti_noninteracting_2023,  nagarajan_espresso_2024, el-badry_red_2023, tanikawa_2023, gaia_collaboration_discovery_2024} and a population of 21 dormant NSs (``Gaia NSs'') \citep{el-badry_19_2024, el-badry_population_2024} in au-scale orbits.

Isolated binary evolution models struggle to explain both BH LMXBs and Gaia compact object binaries. Mass transfer between a massive BH or NS progenitor and a low-mass star is expected to be unstable \citep[e.g.,][]{ivanova_review_2013, ge_adiabatic_2015}. Binary evolution predicts that the orbital energy of a low-mass star is usually insufficient to unbind the envelope of a BH or NS progenitor \citep[e.g.,][]{portegies_zwart_formation_1997, podsiadlowski_formation_2003}, so such systems should merge before the formation of the compact object. While it is possible that some binaries with a highly unequal mass ratio can survive common envelope evolution (CEE), the orbits of surviving post-CEE binaries are predicted to be tight ($P_{\text{orb}} \lesssim 1$ d) \citep[e.g.,][]{webbink_formalism_1984, podsiadlowski_formation_2003, ivanova_review_2013}. Such an evolutionary scenario may apply to BH LMXBs, but it cannot explain the formation of Gaia BHs and NSs, which have orbital periods of years. To make matters worse, natal kicks arising from a supernova can unbind many BH or NS binaries altogether \citep[e.g.,][]{brandt_effects_1995, kotko_2024}. 

Several ideas have been proposed in the literature to form BH LMXBs, including evolution of an intermediate-mass star into the present-day low-mass companion \citep[e.g.,][]{podsiadlowski_intermediate_2000, justham_magnetic_2006}, introduction of additional sources of energy during the common-envelope phase \citep[e.g.,][]{podsiadlowski_explosive_2010, ivanova_recombination_2015}, or dynamical exchanges in a globular cluster followed by binary hardening and eventual mass transfer \citep[e.g.,][]{hills_globular_1976, kremer_accreting_2018}. Various theoretical explanations have also been suggested for the formation of Gaia BHs and NSs, such as modifications to stellar evolution that prevent the BH or NS progenitor from expanding to red supergiant dimensions \citep[e.g.,][]{kruckow_2024, gilkis_2024, li_2024, iorio_2024}, or dynamical interactions in open or globular clusters that produce wide orbits and moderate eccentricities \citep[e.g.,][]{rastello_2023, tanikawa_2024, di_carlo_young_2024,  pina_2024, fantoccoli_2025}.

A promising class of models for forming BH LMXBs and Gaia compact object binaries rely on hierarchical triple evolution \citep[e.g.,][]{naoz_2016}. For example, a BH and low-mass star could have originated in a wide ($\sim 100$ au) orbit with a distant ($\gtrsim 1000$ au) tertiary. After formation of the BH, (eccentric) Kozai-Lidov oscillations could tighten the inner binary, producing a BH LMXB \citep[e.g.,][]{naoz_2016, shariat_cygni_2025}. Another idea is that Kozai-Lidov oscillations induced by an outer tertiary star could delay the onset of common-envelope evolution \citep{generozov_perets_2024}. This could lead to wider initial and final orbital separations, characteristic of Gaia BHs or NSs \citep{generozov_perets_2024}. In such scenarios, compact object binaries could retain faint low-mass star or white dwarf tertiaries at separations of $\sim 100$--$1000$ au, especially if these systems experience weak natal kicks.

In support of this hypothesis, \citet{burdge_nature_2024} recently discovered that V404 Cyg, one of the best-studied BH LMXBs \citep[e.g.,][]{casares_cygni_1992}, is in a hierarchical triple. V404 Cyg contains a $\sim9\,M_{\odot}$ BH \citep{khargharia_cyg_2010} in a 6.4-day orbit with a $\approx0.7\,M_\odot$ star \citep{shahbaz_cygni_1994}. {\it Gaia} astrometry and follow-up spectroscopy revealed that an evolved $1.2\,M_{\odot}$ star $1.4$ arcsec away --- long assumed to be a chance alignment --- is in fact a gravitationally bound tertiary, with a separation of $\approx3500$ au \citep{burdge_nature_2024}. Because the tertiary is weakly bound, with an orbital velocity of only $\sim 1\,\rm km\,s^{-1}$, the BH must have formed with a negligible kick of $< 5$ km s$^{-1}$ \citep{burdge_nature_2024, shariat_cygni_2025}. This discovery corroborates the growing picture that some BHs in binaries were born with a strong kick, while others received virtually no kick at all \citep[e.g.,][]{fragos_understanding_2009, atri_potential_2019, zhao_evidence_2023, vigna-gomez_constraints_2024, burrows_theory_2024, nagarajan_mixed_2024}.

As \citet{burdge_nature_2024} point out, V404 Cyg is nearer and brighter in the optical than most other BH LMXBs --- if it were $\sim$50\% more distant, the tertiary would have been blended with the inner binary, and {\it Gaia} would not have been able to measure its proper motion. Furthermore, if the tertiary were $\gtrsim 10\%$ more massive, it would already be a faint white dwarf below {\it Gaia}’s detection limit; on the other hand, if it were $\gtrsim 20\%$ less massive, it would have been too faint for {\it Gaia} to have measured a proper motion precise enough to confidently establish the physical association \citep{burdge_nature_2024}. This suggests that harder-to-detect tertiaries may be hiding around other known BHs, and that deeper searches could reveal their presence and constrain the hierarchical triple formation channel. 

Motivated by the possibility that there could be undetected tertiaries orbiting both BH LMXBs and Gaia BH/NS binaries, we carried out a near-infrared adaptive optics (AO) imaging campaign to detect them. Since quiescent BH LMXBs can be faint or obscured in the optical, near-infrared high-contrast imaging is ideally suited for this task. Indeed, such techniques have already been used to estimate the companion frequency of nearby high-mass X-ray binaries (HMXBs; \citealt{prasow_2024}). 

The remainder of this paper is organized as follows. In Section~\ref{sec:data}, we describe our campaign to acquire near-infrared images of quiescent BH LMXBs, Gaia BH1, and Gaia NSs with NIRC2 on Keck, aided by the laser guide star (LGS) AO system. In Section~\ref{sec:results}, we use high contrast imaging techniques to place constraints on putative tertiaries to the BH and NS binaries in our sample. In Section~\ref{sec:discussion}, we discuss potential caveats and implications for the formation channels of these binaries. Finally, in Section~\ref{sec:conclusion}, we outline our conclusions and suggest opportunities for further follow-up.

\section{Data}
\label{sec:data}

\subsection{Sample}

% Convert K-magnitudes from AB to Vega by subtracting 1.85 (Blanton et al. 2005)
\begin{deluxetable*}{cccccccc}
\tablecaption{Log of Keck/NIRC2 observations of BH and NS binaries. All targets were observed once in the $K'$ band using the \texttt{bxy3} dither pattern, with $t_{\text{int}}$ referring to the total integration time. Reported $K_s$-band apparent magnitudes (in the Vega system) are taken from the 2MASS catalog \citep{skrutskie_2006} unless otherwise noted. \label{tab:log}}
\tablehead{\colhead{UT Date} & \colhead{Target} & \colhead{\textit{Gaia} DR3 ID} & \colhead{Distance (kpc)} & \colhead{Classification} & \colhead{$K_s$} & \colhead{$t_{\text{int}}$ (s)} & \colhead{Camera} \\
\colhead{(1)} & \colhead{(2)} & \colhead{(3)} & \colhead{(4)} & \colhead{(5)} & \colhead{(6)} & \colhead{(7)} & \colhead{(8)}}
\startdata
2024 August 8 & Gaia NS1 & 6328149636482597888 & $0.732 \pm 0.006$$^{1}$ & Dormant NS & $11.83 \pm 0.03$  & 900 & Narrow \\ 
2024 August 8 & Swift J1727.8-1613 & 4136619720186379264 & $3.4 \pm 0.3$$^{2}$ & BH LMXB & $17.0 \pm 0.2$\tablenotemark{a} & 1800 & Narrow \\ 
2024 August 8 & Gaia BH1 & 4373465352415301632 & $0.477 \pm 0.004$$^{3}$ & Dormant BH & $11.78 \pm 0.02$ & 900 & Narrow \\
2024 August 8 & V4641 Sgr & 4053096388919082368 & $4.7_{-0.6}^{+0.8^{4}}$ & BH LMXB & $12.27 \pm 0.03$ & 900 & Narrow \\
2024 August 8 & MAXI J1820+070 & 4477902563164690816 & $2.96 \pm 0.33$$^{5}$ & BH LMXB & $15.1 \pm 0.1$ & 1800 & Narrow \\
2024 August 8 & J2145+2837 & 1801110822095134848 & $0.242 \pm 0.001$$^{1}$ & Dormant NS & $10.70 \pm 0.02$ & 900 & Narrow \\
2024 August 8 & J2244-2236 & 2397135910639986304 & $0.481 \pm 0.004$$^{1}$ & Dormant NS & $12.11 \pm 0.02$ & 900 & Narrow \\
2024 August 8 & J0036-0932 & 2426116249713980416 & $0.602 \pm 0.007$$^{1}$ & Dormant NS & $11.66 \pm 0.02$ & 900 & Narrow \\
2024 August 8 & J0152-2049 & 5136025521527939072 & $0.408 \pm 0.003$$^{1}$ & Dormant NS & $10.90 \pm 0.03$ & 900 & Narrow \\
2024 August 8 & GRO J0422+32 & 172650748928103552 & $2.49 \pm 0.30$$^{6}$ & BH LMXB & $17.44 \pm 0.09$\tablenotemark{b} & 1800 & Narrow \\
2025 January 29 & J1048+6547 & 1058875159778407808 & $1.092 \pm 0.019$$^{1}$ & Dormant NS & $13.31 \pm 0.04$ & 900 & Narrow \\
2025 April 5 & J0824+5254 & 1028887114002082432 & $0.609 \pm 0.006$$^{1}$ & Dormant NS & $12.25 \pm 0.02$ & 180 & Wide \\
2025 April 5 & J1150-2203 & 3494029910469026432 & $0.575 \pm 0.005$$^{1}$ & Dormant NS & $11.30 \pm 0.03$ & 180 & Wide \\
2025 April 5 & J1449+6919 & 1694708646628402048 & $0.552 \pm 0.003$$^{1}$ & Dormant NS & $11.84 \pm 0.02$ & 180 & Wide \\
2025 April 5 & J1733+5808 & 1434445448240677376 & $0.689 \pm 0.005$$^{1}$ & Dormant NS & $12.33 \pm 0.03$ & 180 & Wide \\
2025 April 5 & J1739+4502 & 1350295047363872512 & $0.888 \pm 0.010$$^{1}$ & Dormant NS & $12.25 \pm 0.02$ & 180 & Wide \\
2025 April 5 & GS 2000+25 & 1834737637190059392 & $2.7 \pm 0.7$$^{7}$ & BH LMXB & $17.0 \pm 0.2$\tablenotemark{c} & 450 & Wide \\
2025 April 5 & J2102+3703 & 1871419337958702720 & $0.657 \pm 0.006$$^{1}$ & Dormant NS & $12.47 \pm 0.03$ & 180 & Wide \\
\enddata
\tablenotetext{a}{Estimated based on \citet{pecaut_2013} using the spectral type classification of K4($\pm1$)V from \citet{matasanchez_2025}.}
\tablenotetext{b}{Mean magnitude in quiescence measured by \citet{gelino_2003}.}
\tablenotetext{c}{Mean magnitude in quiescence measured by \citet{beekman_gs_1996}.}
\tablerefs{[1] \citet{el_badry_2024}, [2] \citet{matasanchez_2025}, [3] \citet{el-badry_sun-like_2023}, [4] \citet{bailerjones_2021}, [5] \citet{atri_2020}, [6] \citet{gelino_2003}, [7] \citet{jonker_2004}.}
\end{deluxetable*}

We list the targets in our sample, along with their literature $K_s$-band magnitudes, in Table~\ref{tab:log}. In total, we observed five quiescent BH LMXBs, Gaia BH1, and twelve Gaia NSs. This represents approximately $25\%$ of the dynamically confirmed BH LMXB population, $33\%$ of the known Gaia BH population, and $57\%$ of the known Gaia NS population. We selected our targets based on their observability during the months of our Keck/NIRC2 campaign. In particular, we were able to follow up a majority of the targets that were observable from the Northern Hemisphere in August 2024 and April 2025.\footnote{While we did not observe Gaia BH3 during our campaign, that binary was observed with VLTI/GRAVITY by \citet{kervella_vlti_2025}. The primary goal of their observations was to detect emission from the BH due to wind accretion. No emission was detected a contrast level of 6.8 mag in the $K$-band.} Though the sample could be made more complete with future NIRC2 imaging, eight dynamically confirmed BH LMXBs, Gaia BH2, and five Gaia NSs are located at $\delta < -40^{\circ}$ and are therefore inaccessible from Keck Observatory.

\subsection{Observations}

We used NIRC2 (PI: K.\ Matthews), a near-infrared camera on the Keck II telescope, with the Keck laser guide star (LGS) AO system \citep{wizinowich_lgs_2006} to pursue diffraction-limited imaging of our targets. Unlike natural guide star AO, the LGS system can operate with tip-tilt stars as faint as $V = 18$ mag, substantially increasing sky coverage. 

\begin{figure*}
    \centering
    \includegraphics[width=0.75\textwidth]{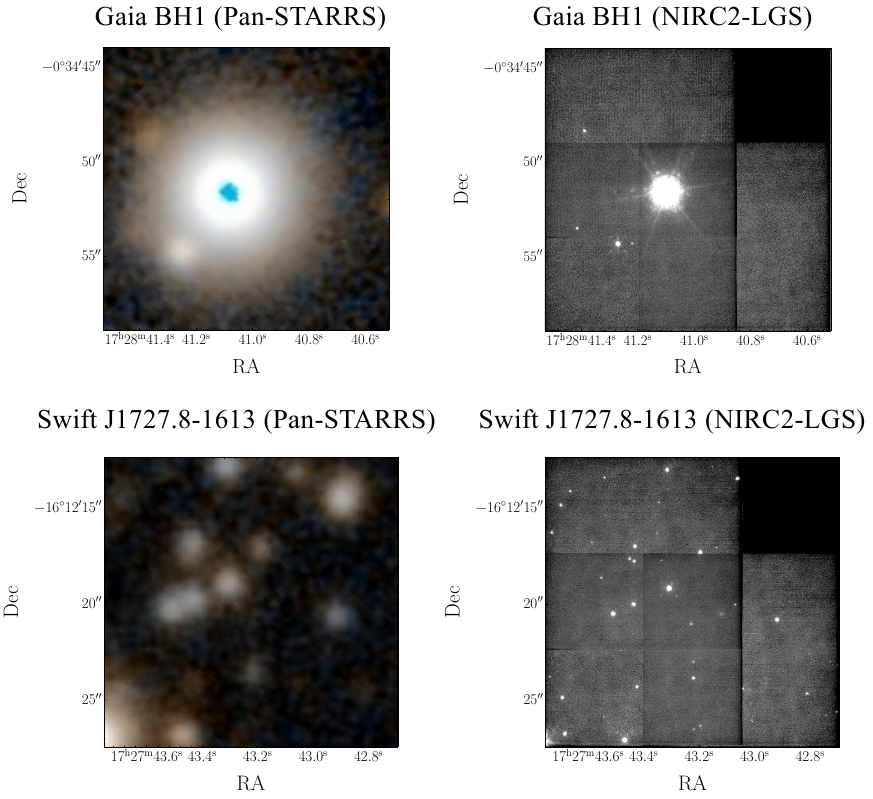}
    \caption{Comparison of Pan-STARRS \citep{PanSTARRS} optical survey images and NIRC2-LGS images (this work) for Gaia BH1 and Swift J1727. The targets are identified with purple cross-hairs in the Pan-STARRS images. Artifacts due to persistence have been masked (see Section~\ref{sec:persistence}). Due to improved depth and spatial resolution, we detect numerous neighboring sources that were unknown to previous survey imaging.}
    \label{fig:comparison}
\end{figure*}

During the first half of our campaign, in which we observed most of the quiescent BH LMXBs, we used the $10\arcsec\times10\arcsec$ narrow camera (which has a pixel scale of 0.009942 arcsec/pixel) to maximize spatial resolution. During the second half, in which we observed most of the Gaia NSs, we used the $40\arcsec\times40\arcsec$ wide camera (which has a pixel scale of 0.039686 arcsec/pixel) to maximize our field of view and facilitate photometric calibration with a larger set of reference stars. We observed in the $K'$ filter ($\approx$1.9--2.3~$\mu$m), which offers lower thermal background than the standard $K$ filter and thus improved depth for a given integration time \citep{wainscoat_cowie_1992}. The bandpass of the $K'$ filter is sufficiently similar to that of the 2MASS $K_s$ filter to permit direct comparison to published catalog magnitudes.

We acquired several co-adds at each of three dither positions taken in the \texttt{bxy3} pattern, with total integration times listed for each target in Table~\ref{tab:log}. Images were reduced and stacked using the KAI data reduction pipeline\footnote{\texttt{github.com/Keck-DataReductionPipelines/KAI/tree/dev}} \citep{lu_kai_2021}, which performs dark subtraction, flat-field corrections, hot pixel and cosmic ray removal, and image alignment. The pipeline also accounts for the NIRC2 astrometric distortion solution \citep{service_lu_2016}.

\begin{figure*}
    \centering
    \includegraphics[width=\textwidth]{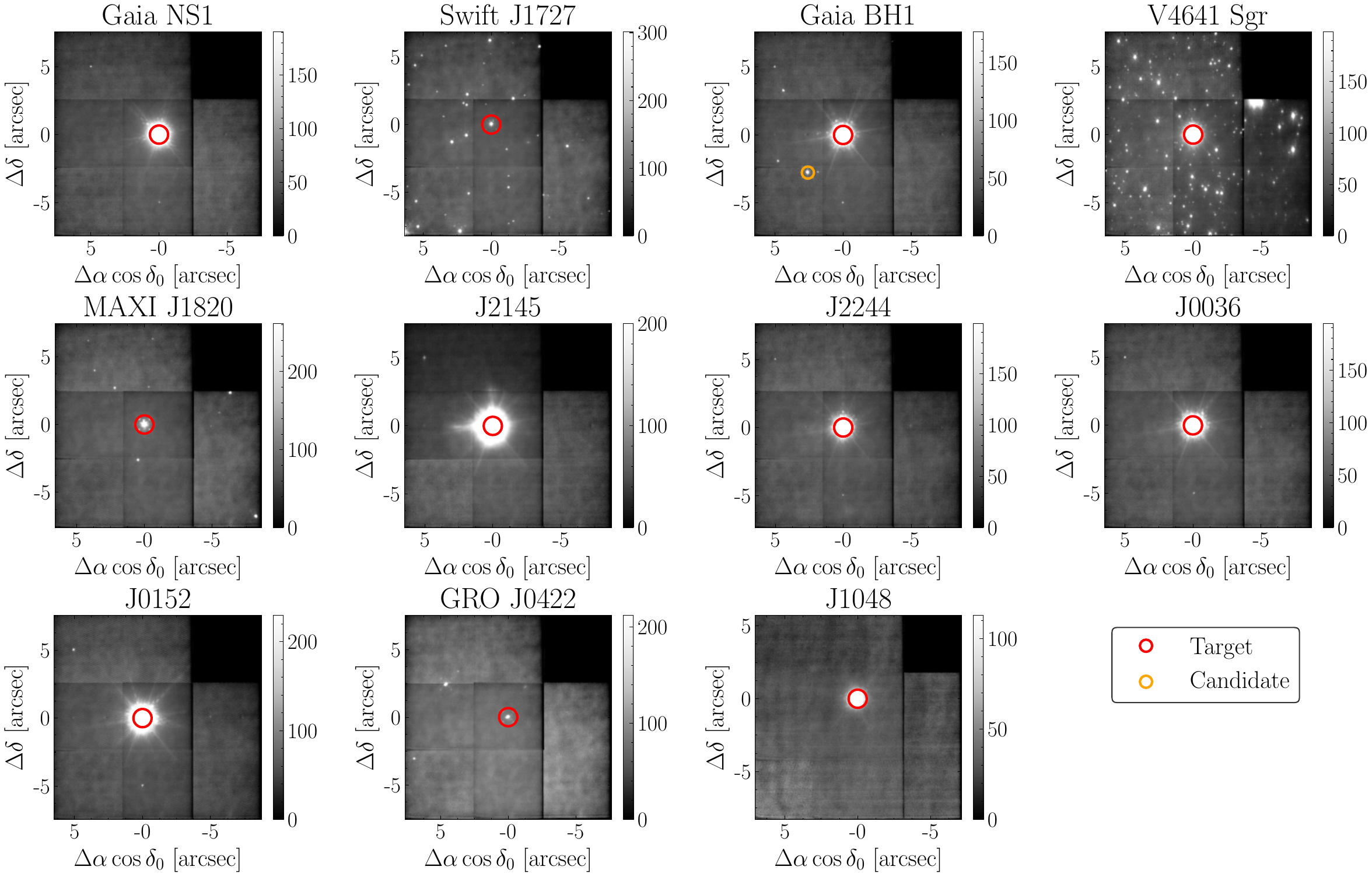}
    \caption{Reduced and stacked NIRC2-LGS images taken with the narrow ($10\arcsec \times 10\arcsec$) camera. The location of each target is marked with a red circle. We resolve numerous sources previously undetected in survey imaging, especially for dense fields at low Galactic latitudes. Detected SNR $> 5$ point sources with low (i.e., $< 0.05$) chance alignment probabilities, which we consider to be tertiary candidates (Table~\ref{tab:ao_cands}), are marked with orange circles.}
    \label{fig:narrow_images}
\end{figure*}

\begin{figure*}
    \centering
    \includegraphics[width=\textwidth]{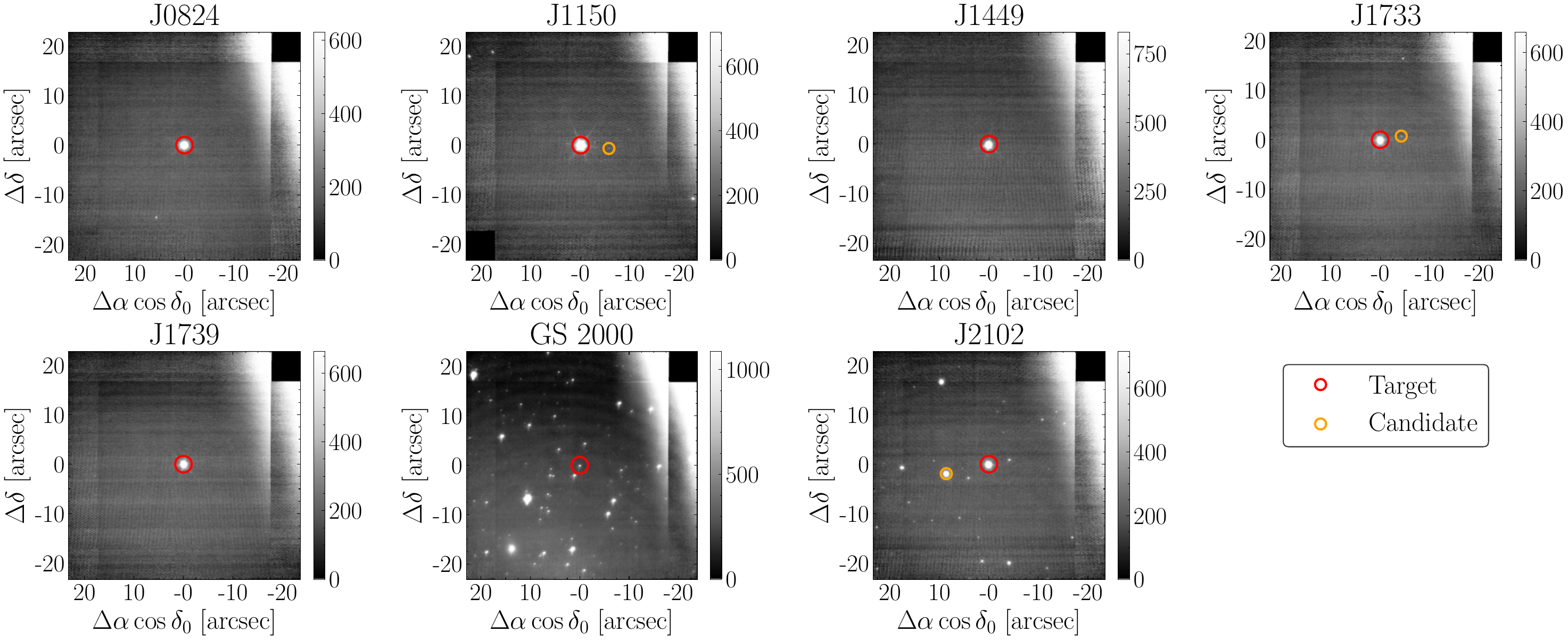}
    \caption{Same as Figure~\ref{fig:narrow_images}, but for reduced and stacked images taken with the wide ($40\arcsec \times 40\arcsec$) camera.}
    \label{fig:wide_images}
\end{figure*}

To showcase the improved depth and spatial resolution of our campaign relative to archival imaging, we compare our reduced images of Gaia BH1 and Swift J1727 to cutouts from Pan-STARRS \citep{PanSTARRS} in Figure~\ref{fig:comparison}. Several faint point sources visible in our NIRC2-LGS images are either blended with the target or unresolved in the Pan-STARRS cutouts, underscoring the need for AO-assisted imaging in searches for tertiary companions. We present all of our reduced and stacked NIRC2-LGS images taken with the narrow and wide cameras in Figures~\ref{fig:narrow_images} and \ref{fig:wide_images}, respectively. We detect numerous nearby sources that were unknown to previous survey imaging, especially in crowded fields (e.g., Swift J1727 and V4641 Sgr). On the other hand, our images of sparse fields at high Galactic latitudes (e.g., for the Gaia NSs) reveal few new sources, despite the higher spatial resolution relative to survey imaging. We describe how we identified genuine sources with signal-to-noise ratio (SNR) $> 5$ in each field-of-view in Section~\ref{sec:photometry}, consider their chance alignment probabilities in Section~\ref{sec:chance_alignments}, and search for stellar companions at close separations in Section~\ref{sec:rdi}.

\subsubsection{Persistence}
\label{sec:persistence}

Several of our bright targets in Figure~\ref{fig:narrow_images} (i.e., Gaia NS1, Gaia BH1, J2244, J0036, and J0152) have apparent companions at $\sim 5\arcsec$ separations to the south and northeast. These appear at almost the same position on the detector for each source and are artifacts that arise from persistence between subsequent dithers. For long exposures, signal from a bright star can remain on the detection array even after the source is removed from the field-of-view, decaying at a rate inversely proportional to time \citep{nirc2_persistence_2005}. The Aladdin III InSb detector on NIRC2 is known to be susceptible to this effect \citep{nirc2_persistence_2005}. We identified persistence features by visual comparison of subsequent dithers, and removed any associated artifacts from the source lists constructed in Section~\ref{sec:photometry}.

\section{Results}
\label{sec:results}

\subsection{Photometry}
\label{sec:photometry}

We used 2D Gaussian fitting to estimate the full-width half-maximum (FWHM) of the AO-corrected point spread function (PSF) in each of our stacked images. Then, we used the DAOStarFinder algorithm from the \texttt{photutils} package to identify sources consistent with this PSF in each field of view, requiring signal-to-noise ratio (SNR) $>5$.\footnote{To identify partially resolved double stars, \textit{Gaia} DR3 introduced \texttt{ipd\_frac\_multi\_peak}, which represents the percentage of CCD observations where image parameter determination detected more than one peak. Only sources with \texttt{ipd\_frac\_multi\_peak} $< 2$ received astrometric binary solutions in DR3 \citep{halbwachs_gaia_2023}. Hence, any tertiary resolved by NIRC2 for the Gaia compact objects in our sample would have to be faint enough to satisfy that criterion.} The typical PSF FWHM in our images was $\approx 0.12\arcsec$, corresponding to 12 pixels and 3 pixels for the narrow camera and wide camera, respectively. For comparison, the diffraction limit of the Keck II 10m telescope in the $K'$ band is $\approx 0.05\arcsec$. Our measured FWHM indicates that, while AO corrections significantly improved upon seeing-limited resolutions, they did not reach the diffraction limit.

We estimated the $K_s$-band apparent magnitudes of these sources by calibrating to the 2MASS catalog \citep{skrutskie_2006}. In detail, we computed the flux of each target and each source within an aperture of radius equivalent to the FWHM of the image. We estimated and subtracted the local background using an annulus with inner and outer radius $1.5\times$ and $3\times$ the size of the aperture, respectively. Then, using the known $K_s$-band magnitudes of our targets, we inferred the apparent $K_s$-band magnitudes of the sources, for which we adopt a conservative uncertainty of $\pm 0.5$ mag.

\subsection{Chance Alignment Probabilities}
\label{sec:chance_alignments}

% For the closest resolved neighbor of flux $F_{\text{thresh}}$ at separation $r$, we calculate its chance alignment probability as:

% \begin{equation}
%     P(F_{\text{thresh}}, r) = 1 - e^{-p(F_{\text{thresh}}, r)},
% \end{equation}

% where $p(F_{\text{thresh}}, r)$ is the probability of finding a source at least as bright as flux threshold $F_{\text{thresh}}$ at that separation:

% \begin{equation}
%     p(F_{\text{thresh}}, r) = n(F \geq F_{\text{thresh}}) \frac{\pi r^2}{A}
% \end{equation}

% Here, $A$ is the total field of view, and $n(F \geq F_{\text{thresh}}$ is the total number of 5$\sigma$ sources detected by DAOStarFinder. 

While we resolve numerous stars that were previously undetected in survey imaging, many of these are background or foreground sources. Formally, for a point source of apparent magnitude $K_{s, \text{candidate}}$, we estimate the chance alignment probability $P$ as:

\begin{equation}
    P(\Sigma, \Theta) = 1 - e^{-\pi \Sigma \Theta^2},
\end{equation}

\noindent where $\Theta$ is the projected angular separation and $\Sigma = \Sigma(K_s < K_{s, \text{candidate}})$ is the local surface density of physically unassociated sources that are at least as bright as the candidate \citep[e.g.,][]{correia_survey_2006}. To estimate the surface density in each field, we used the Galactic stellar population model TRILEGAL \citep{girardi_trilegal_2005} to predict the star counts as a function of apparent $K_s$-band magnitude in a 1 square degree region around the location of each target. For V4641 Sgr, which is located in a very dense field, we used a total field area of 1 square arcmin instead. 

We consider sources with $P(\Sigma, \Theta) < 0.05$ as tertiary candidates. We identify these sources with orange circles in Figures~\ref{fig:narrow_images} and \ref{fig:wide_images}. Sources with $P > 0.05$ could still be tertiaries, but lack strong evidence to rule out a chance alignment. We provide projected angular and physical (i.e., $a_{\text{proj}}$) separations, position angles (PAs), estimated $K_s$-band magnitudes, and chance alignment probabilities for these wide tertiary candidates in Table~\ref{tab:ao_cands}. The absolute magnitudes are computed using distances from the literature \citep{corral-santana_blackcat_2016, el-badry_sun-like_2023, el-badry_19_2024, el-badry_population_2024} and extinctions from the all-sky 3D dust map of \citet{wang_dust_2025}, assuming that the candidates are at the same distance as the target. Of the four candidates, we determine three to be unassociated with the targets because they have inconsistent parallax and/or proper motion measurements in DR3. The remaining candidate (located $4.3\arcsec$ to the west of J1733 and undetected in DR3) remains viable. Using the main sequence mass-magnitude relation of \citet{pecaut_2013}, we estimate the spectral type and mass of each tertiary candidate, and report them in Table~\ref{tab:ao_cands}. We find that the source next to J1733 is predicted to be an $\approx 0.2\,M_{\odot}$ M4V dwarf at a separation of $\sim 3000$ au. This candidate requires further follow-up to measure its proper motions and test its physical association.

\begin{deluxetable*}{cccccccccc}
\tablecaption{Position angles, separations, and magnitudes for tertiary candidates at wide separations, detected at SNR $>5$ in high-resolution, near-infrared adaptive optics imaging. Each of these candidates has chance alignment probability $\lesssim 5\%$. Absolute magnitudes are estimated by assuming that the candidate is at the same distance as the target. We then estimate spectral types and masses based on the main sequence relation of \citet{pecaut_2013}. \textit{Gaia} DR3 astrometry reveals that three of these sources are unassociated with their respective targets. The remaining candidate, located $\approx 4.3\arcsec$ to the west of J1733, is undetected in DR3 and remains viable. \label{tab:ao_cands}}
\tablehead{\colhead{Target} & \colhead{PA} & \colhead{$\Theta$} & \colhead{$a_{\text{proj}}$} & \colhead{$m_{K_s}$} & \colhead{$M_{K_s}$} & \colhead{$P(\Sigma, \Theta)$}  & \colhead{Est.\ Sp.\ Type} & \colhead{Est.\ Mass} & \colhead{Classification} \\
& $^{\circ}$ & $\arcsec$ & au & $\pm 0.5$ mag & & & & $M_{\odot}$ & \\
\colhead{(1)} & \colhead{(2)} & \colhead{(3)} & \colhead{(4)} & \colhead{(5)} & \colhead{(6)} & \colhead{(7)} & \colhead{(8)} & \colhead{(9)} & \colhead{(10)}}
\startdata
Gaia BH1 & 136.880 & 3.785 & 1806 & 16.2 & 7.8 & 0.038 & M4.5V & 0.184 & Unassociated (DR3) \\
J1150 & 263.322 & 5.755 & 3309 & 16.7 & 7.9 & 0.023 & M4.5V & 0.184 & Unassociated (DR3) \\
J1733 & 279.909 & 4.312 & 2971 & 16.8 & 7.6 & 0.016 & M4V & 0.23 & Candidate \\
J2102 & 98.375 & 8.809 & 5787 & 12.6 & 3.5 & 0.045 & G8V & 0.94 & Unassociated (DR3) \\
\enddata
\end{deluxetable*}

\subsection{Reference Star Differential Imaging}
\label{sec:rdi}

To search for companions at close ($\lesssim 1\arcsec$) separations, post-processing is required to suppress the contribution of the target PSF and achieve high contrast sensitivity. To model the AO-corrected PSF, we used the \texttt{VIP} package \citep{gomez_vip_2017, 2023JOSS....8.4774C} to perform reference star differential imaging (RDI) with full-frame principal component analysis (PCA) using the Karhunen-Lo\'eve Image Processing (KLIP) algorithm \citep{soummer_klip_2012}. We constructed libraries of reference images for both the narrow and wide cameras by aligning and cropping individual dithers taken over the course of our NIRC2-LGS campaign (i.e., across all targets and science programs). To select the optimal number of principal components (PCs) for our analysis, we injected faint, fake companions at multiple position angles and radii within the speckle-dominated regime ($\lesssim 1\arcsec$) of each image. We then ran several PCA reductions, varying the number of PCs between 1 and the size of the corresponding reference library. We measured the recovered signal-to-noise ratio (SNR) as a function of the number of PCs, adopting the median value at which this metric plateaued as our optimal choice. The plateau, which represents the trade-off between speckle suppression and companion over-subtraction, was found to occur at a similar number of PCs across targets for a given camera. This suggests that it is primarily set by the PSF sampling and reference library properties rather than individual target characteristics.  Indeed, the wide camera has a larger pixel size than the narrow camera, which leads to the PSF being less well sampled and corresponds to a smaller optimal number of PCs.

\begin{figure*}
    \centering
    \includegraphics[width=\textwidth]{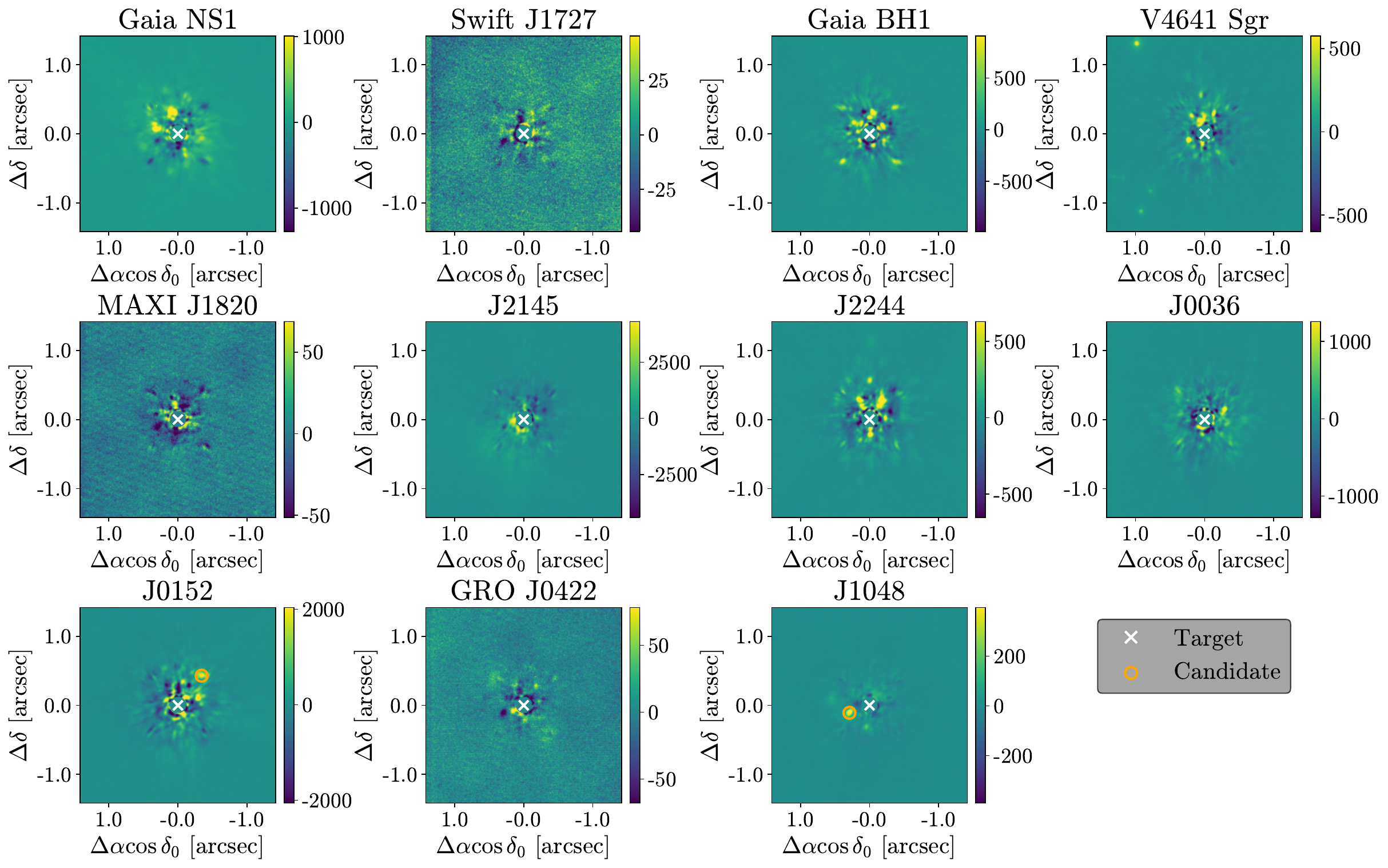}
    \caption{High contrast narrow camera images of BH and NS binaries with low-mass companions, following post-processing with principal component analysis to model and subtract the target PSFs. Specifically, we perform reference star differential imaging with 16 principal components. We mask the central $0.12\arcsec$, where residuals are strongest. Detected $5\sigma$ candidates with chance alignment probability $< 0.05$ are marked with orange circles. We stretch the images to emphasize the detected candidates (the central speckle regions are saturated as a result).}
    \label{fig:narrow_rdi}
\end{figure*}

\begin{figure*}
    \centering
    \includegraphics[width=\textwidth]{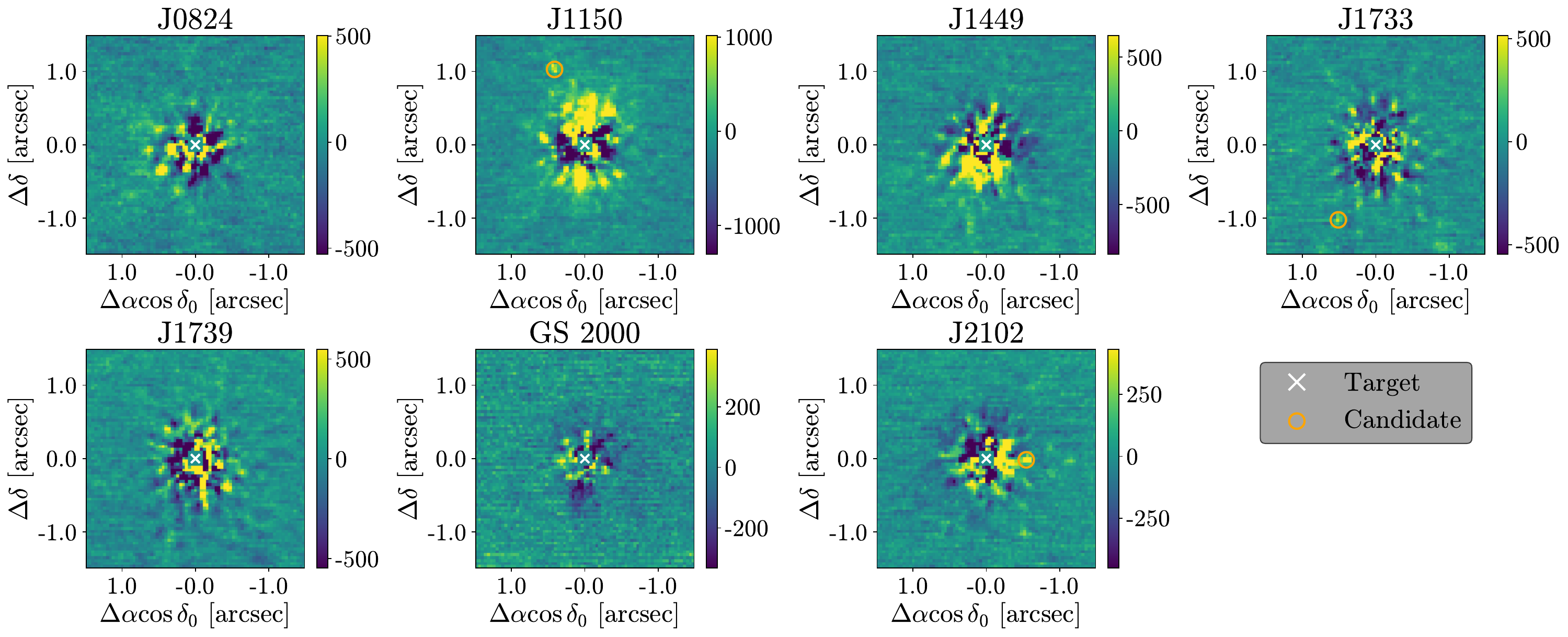}
    \caption{Same as Figure~\ref{fig:narrow_rdi}, but for targets observed with the wide camera. In this case, we use reference star differential imaging with 12 principal components to model and subtract the target PSFs.}
    \label{fig:wide_rdi}
\end{figure*}

We adopted 16 PCs and 12 PCs for the modeling of the target PSFs in images taken with the narrow and wide cameras, respectively. To verify the robustness of our source detection to this choice, we repeated our analysis using different numbers of PCs near our adopted values, finding that we consistently recovered the candidates listed in Table~\ref{tab:pca_cands} (see below). We provide the post-processed high-contrast images taken with the narrow and wide cameras in Figures~\ref{fig:narrow_rdi} and \ref{fig:wide_rdi}. We mask the central $0.12\arcsec$ (i.e., $\approx 1\times$ FWHM or $2\times$ diffraction limit) region of each image, where residuals are strongest.

We generated signal-to-noise (SNR) maps with the \texttt{VIP} package \citep{gomez_vip_2017, 2023JOSS....8.4774C} to identify 5$\sigma$ detections in our PSF-subtracted images. Specifically, we adopt the definition of SNR given in \citet{mawet_fundamental_2014}, which uses a two-sample \emph{t}-test to compare the flux of a candidate against the flux of background resolution elements at the same radial separation and accounts for the effect of small-number statistics at small angles. SNR is then converted into Gaussian significance by matching the false positive fractions predicted by the normal distribution and the appropriate Student's \emph{t} distribution at that radial separation. For each detected candidate, we derived the best-fit separation, position angle, and flux within a FWHM aperture using the negative fake companion \citep[NEGFC;][]{2010Sci...329...57L, 2010Natur.468.1080M, wertz_negfc_2017} technique with the Nelder-Mead simplex algorithm. We also measured the flux of the corresponding targets within an aperture of the same size.\footnote{When estimating the flux of a faint companion, the NEGFC technique accounts for the average noise level within a FWHM aperture in the speckle-dominated regime. On the other hand, the contribution of the background to the flux of a bright target within an aperture of the same size is negligible.} Using the known 2MASS $K_s$-band magnitudes of each target, we converted these flux measurements to estimated apparent magnitudes. We then computed chance alignment probabilities using TRILEGAL, as described in Section~\ref{sec:chance_alignments}.

We report the projected angular and physical separations, position angles (PAs), magnitudes, and chance alignment probabilities of the $5 \sigma$ sources with chance alignment probabilities $< 0.05$ in Table~\ref{tab:pca_cands}. These close tertiary candidates are marked with orange circles in Figure~\ref{fig:narrow_rdi} and \ref{fig:wide_rdi}. As before, we use literature distances \citep{corral-santana_blackcat_2016, el-badry_sun-like_2023, el-badry_19_2024, el-badry_population_2024} and extinctions from the all-sky 3D dust map of \citet{wang_dust_2025} to compute absolute $K_s$-band magnitudes for these candidates, assuming that they are at the same distance as the targets. Then, using the main sequence mass-magnitude relation of \citet{pecaut_2013}, we estimate spectral types and masses, providing them in Table~\ref{tab:pca_cands} as well.

\begin{deluxetable*}{cccccccccc}
\tablecaption{Position angles, separations, and magnitudes for tertiary candidates at close separations, detected at $5\sigma$ significance in high contrast images. Each of these candidates has chance alignment probability $\lesssim 5\%$. As in Table~\ref{tab:ao_cands}, spectral types and masses are estimated based on the main sequence relation of \citet{pecaut_2013}, assuming that the candidate is at the same distance as the target. While visual inspection and injection-recovery tests suggest that all of these candidates are likely to be artifacts arising from imperfect PSF subtraction (Appendix~\ref{sec:injection_recovery}), we cannot definitively rule out an astrophysical origin. \label{tab:pca_cands}}
\tablehead{\colhead{Target} & \colhead{PA} & \colhead{$\Theta$} & \colhead{$a_{\text{proj}}$} & \colhead{$m_{K_s}$} & \colhead{$M_{K_s}$} & \colhead{$P(\Sigma, \Theta)$}  & \colhead{Est.\ Sp.\ Type} & \colhead{Est.\ Mass} & \colhead{Best Guess} \\
& $^{\circ}$ & $\arcsec$ & au & & & & & $M_{\odot}$ & \\
\colhead{(1)} & \colhead{(2)} & \colhead{(3)} & \colhead{(4)} & \colhead{(5)} & \colhead{(6)} & \colhead{(7)} & \colhead{(8)} & \colhead{(9)} & \colhead{(10)}}
\startdata
J0152 & 318.394 & 0.553 & 226 & 15.2 & 7.1 & 0.00004 & M3.5V & 0.27 & Artifact (Speckle Noise) \\
J1048 & 110.491 & 0.312 & 341 & 16.7 & 6.5 & 0.00004 & M3V & 0.37 & Artifact (Speckle Noise) \\
J1150 & 21.861 & 1.090 & 627 & 18.0 & 9.2 & 0.00164 & M6V & 0.102 & Artifact (Diffraction Spike) \\
J1733 & 153.593 & 1.143 & 788 & 19.7 & 10.5 & 0.00401 & M9.5V & 0.078 & Artifact (Diffraction Spike) \\
J2102 & 268.758 & 0.555 & 365 & 18.6 & 9.5 & 0.00863 & M6.5V & 0.093 & Artifact (Speckle Noise) \\
\enddata
\end{deluxetable*}

Based on visual inspection and injection-recovery tests, we find that all of these candidates could be PSF artifacts arising from radial diffraction spikes or imperfect subtraction of the target PSF (see Appendix \ref{sec:injection_recovery}). Indeed, the speckle noise is expected to be non-Gaussian \citep[e.g.,][]{mawet_fundamental_2014}, in which case spurious 5$\sigma$ detections would not be uncommon. Nevertheless, we cannot definitively reject these sources as tertiary candidates from a single epoch of imaging. Further follow-up with angular differential imaging (ADI) or spectral differential imaging (SDI), which provide independent speckle discrimination, could definitively confirm or rule out these $5\sigma$ detections. For example, a re-detection at a second epoch would establish a candidate as an astrophysical source, and enable a common proper motion test to determine whether the tertiary candidate is gravitationally bound.

\subsection{Contrast Curves}
\label{sec:contrast_curves}

\begin{figure*}
    \centering
    \includegraphics[width=\textwidth]{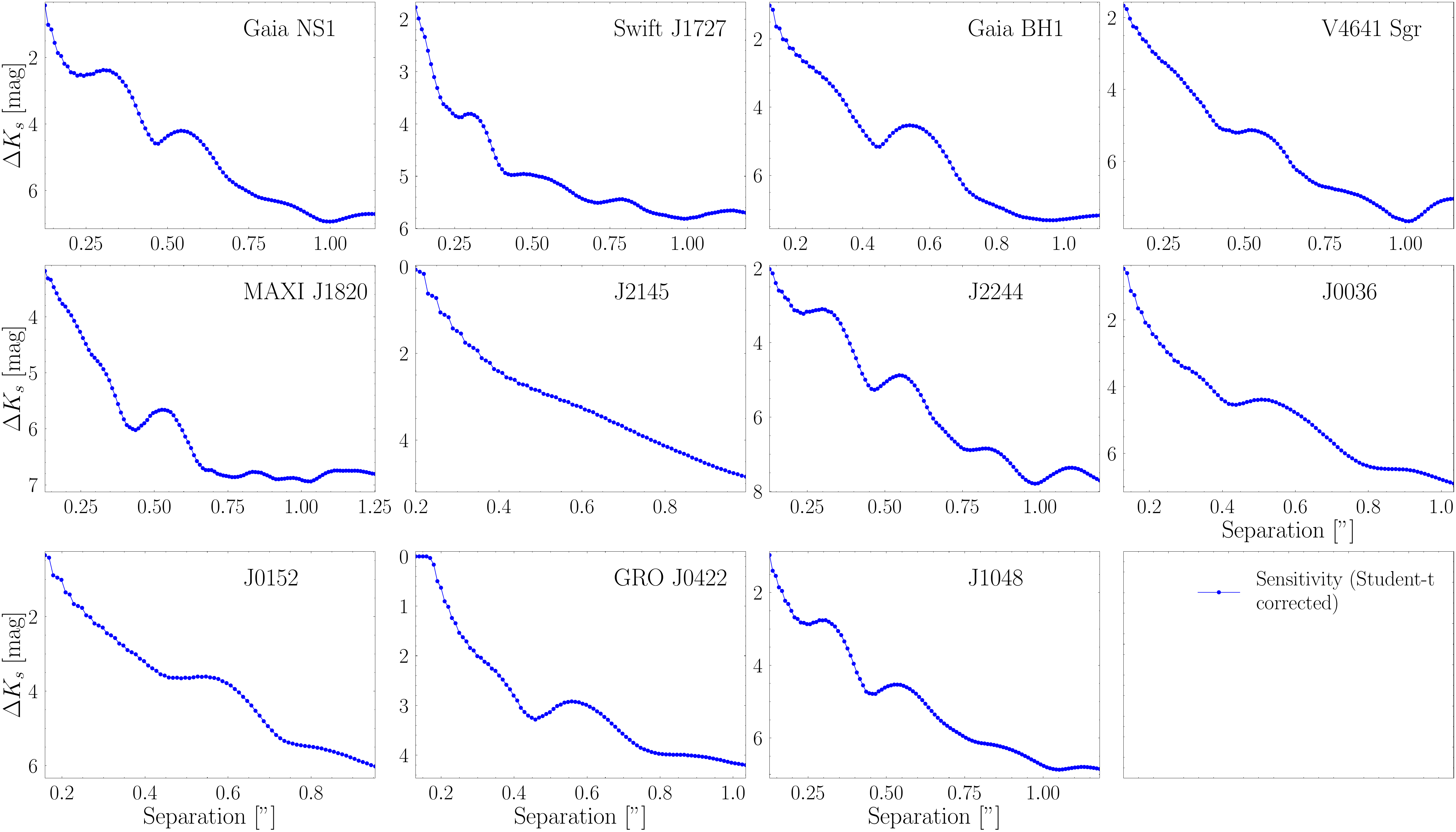}
    \caption{Contrast curves for targets observed with the narrow camera, accounting for both the noise level of the post-processed image and the throughput of the RDI algorithm. The curves represent the 5$\sigma$ detection limit on putative companions as a function of angular separation. We apply a Student-$t$ correction to the result from Gaussian statistics to account for the small number of resolution elements at close separations \citep{mawet_fundamental_2014}.}
    \label{fig:narrow_contrast}
\end{figure*}

\begin{figure*}
    \centering
    \includegraphics[width=\textwidth]{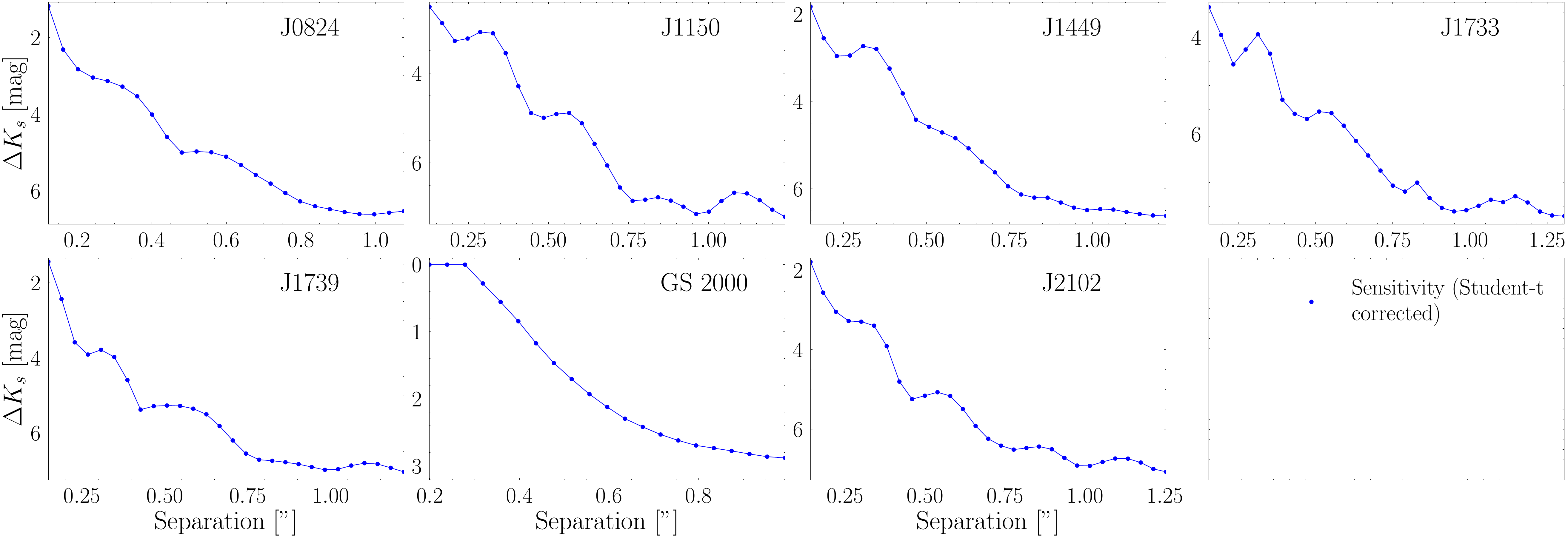}
    \caption{Same as Figure~\ref{fig:narrow_contrast}, but for targets observed with the wide camera.}
    \label{fig:wide_contrast}
\end{figure*}

While we do not detect many promising candidates, tertiaries that fall below our NIRC2-LGS detection threshold may still exist. To place constraints on undetected tertiaries, we use the \texttt{VIP} package \citep{gomez_vip_2017, 2023JOSS....8.4774C} to compute $5\sigma$ contrast curves for each of our targets. Specifically, for each post-processed image, we construct concentric annuli at increasing separations (beginning at a radius of $1$ FWHM). Within each annulus, we place apertures at twelve evenly distributed position angles and estimate the background noise as the standard deviation of the measured aperture fluxes. The aperture diameter is set equal to the measured FWHM. We account for the throughput of the PCA-RDI algorithm by injecting synthetic companions at a range of radii and position angles and measuring their aperture fluxes. Using the estimated target flux from Section~\ref{sec:rdi}, we then compute the 5$\sigma$ detection limit in $\Delta$~mag as a function of separation from the target. Since the noise is expected to be non-Gaussian, we apply a Student-\emph{t} correction to account for small number statistics at close separations \citep[][]{mawet_fundamental_2014}. We show all of our contrast curves in Figures~\ref{fig:narrow_contrast} and \ref{fig:wide_contrast}.

\subsection{Companion Constraints}
\label{sec:companion_constraints}

\begin{figure*}
    \centering
    \includegraphics[width=\textwidth]{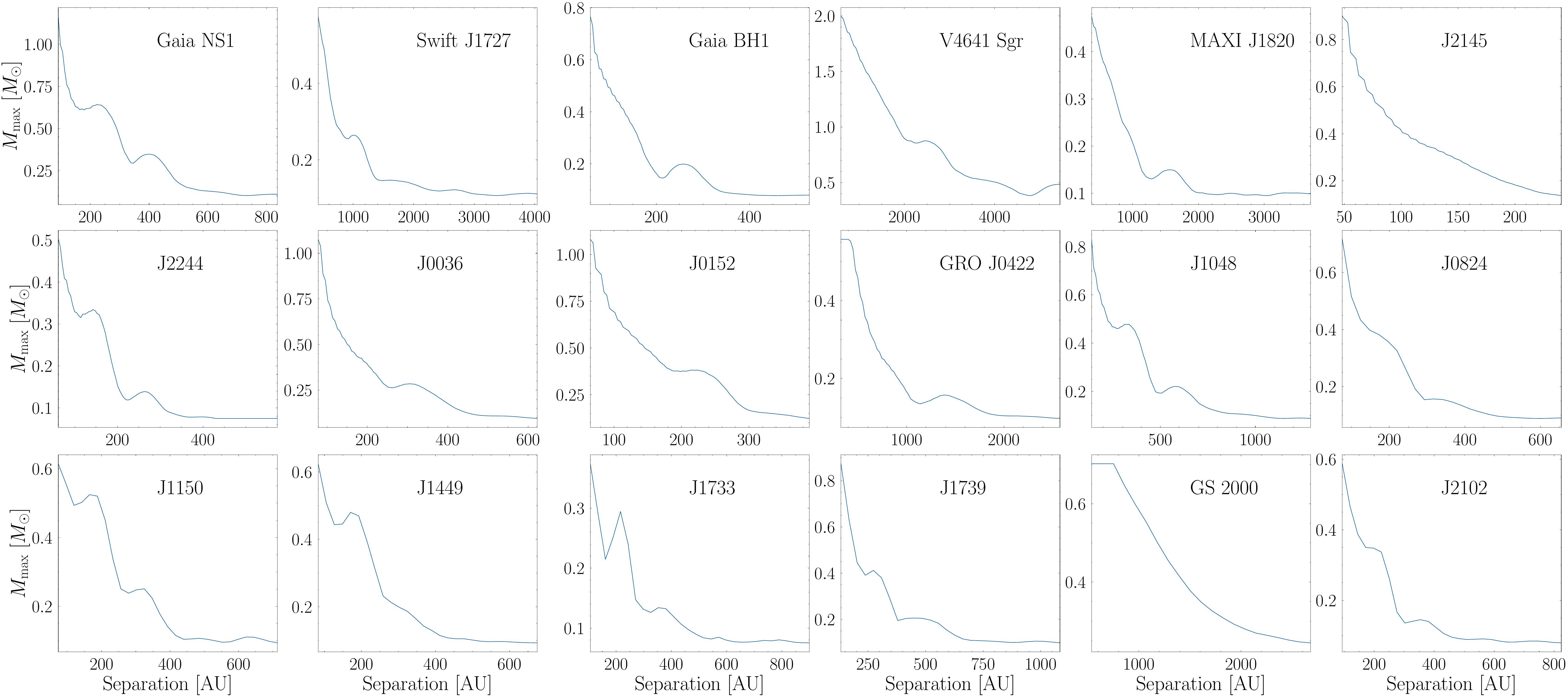}
    \caption{Upper limits on main sequence (MS) companions as a function of projected physical separation. We use $K_s$-band absolute magnitude-mass relations \citep{mann_mdwarf_2019, choi_2016} and our 5$\sigma$ contrast curves to derive limits on the maximum mass of MS tertiaries that are consistent with a non-detection. In general, we can rule out most MS companions above the hydrogen burning limit at separations of $\gtrsim 500$ au for the Gaia compact objects and $\gtrsim 2000$ au for the more distant BH LMXBs.}
    \label{fig:ms_constraints}
\end{figure*}

From the Student-\emph{t}-corrected contrast curves, we infer maximum masses for main sequence (MS) tertiaries and maximum effective temperatures for white dwarf (WD) tertiaries that are consistent with a non-detection as a function of projected separation. In detail, to place constraints on low-mass MS tertiaries, we interpolate between mass and absolute $K_s$-band magnitude using  the M-dwarf relation derived by \citet{mann_mdwarf_2019}. In cases where extrapolation to higher masses is required, we interpolate between mass and absolute $K_s$-band magnitude using a 1.0 Gyr solar-metallicity isochrone from the Mesa Isochrones and Stellar Tracks (MIST) library \citep{choi_2016}.\footnote{The exact age of the isochrone is not critical --- we choose an age where, in the mass range of interest, the least massive tertiaries have reached the main sequence and the most massive tertiaries have not yet left the main sequence.} We then use literature distances to calculate apparent $K_s$-band magnitudes \citep{corral-santana_blackcat_2016, el-badry_sun-like_2023, el-badry_19_2024, el-badry_population_2024}, applying extinctions from the 3D dust map of \citet{wang_dust_2025}. We present our inferred constraints on MS companions in Figure~\ref{fig:ms_constraints}. In general, for the astrometric compact object binaries, we are able to rule out all plausible MS tertiaries above the hydrogen burning limit of $\approx 0.08\,M_{\odot}$ at separations $\gtrsim 500$ au. For the more distant BH LMXBs, we rule out all plausible MS tertiaries at larger separations of $\gtrsim 2000$ au (with the exception of V4641 Sgr, which has a B9III donor \citep{v4641sgr_improved_2014} at a distance of 4.7 kpc \citep{bailer_jones_2021} and thus weaker constraints). We truncate all curves at the hydrogen burning limit, as Gyr-old brown dwarf tertiaries would not be detectable.

\begin{figure*}
    \centering
    \includegraphics[width=\textwidth]{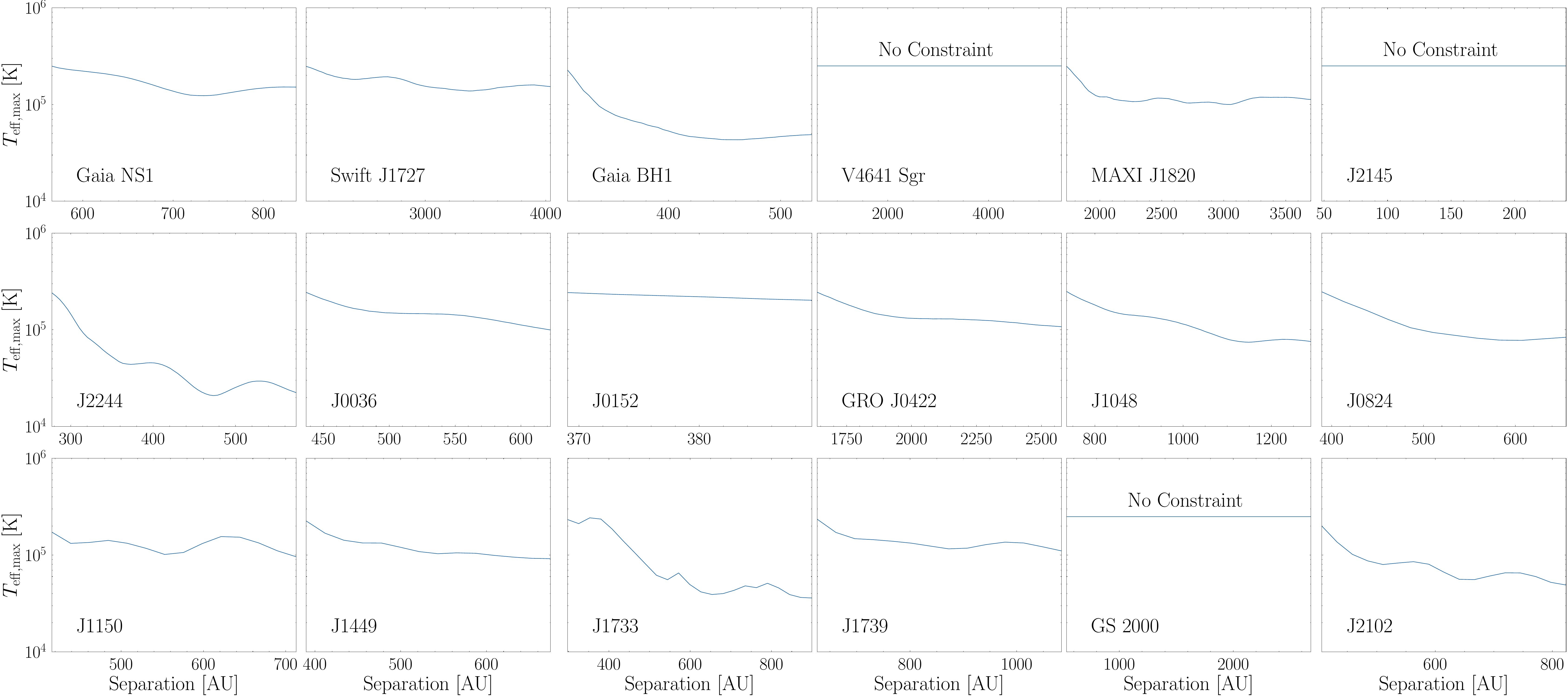}
    \caption{Upper limits on white dwarf (WD) companions as a function of projected physical separation. Modeling the WDs as blackbodies of radius $0.01\,R_{\odot}$ and using our 5$\sigma$ contrast curves, we derive limits on the maximum effective temperature of WD tertiaries that are consistent with a non-detection. In most cases, the observations are consistent with the presence of a WD tertiary across a wide range of $T_{\text{eff}}$.}
    \label{fig:wd_constraints}
\end{figure*}

To place constraints on the effective temperature of WD tertiaries, we assume a blackbody spectrum and a typical radius of $0.01\,R_{\odot}$. We calculate the corresponding absolute $K_s$-band magnitudes using \texttt{pyphot}, which computes synthetic photometry.\footnote{https://mfouesneau.github.io/pyphot/} We convert to apparent $K_s$-band magnitudes using literature distances, applying extinctions based on the 3D dust map of \citet{wang_dust_2025} as before. We present the inferred constraints on WD companions in Figure~\ref{fig:wd_constraints}. We truncate the curves at an upper bound of $250,000$ K, corresponding to the estimated effective temperature of the hottest WD known \citep{werner_hottest_2015}. We find that we can rule out hot, young $\gtrsim 10^5$ K WD tertiaries to the Gaia compact object binaries at projected separations $\gtrsim 500$ au (with the exception of J2145, the closest and brightest object in our sample). On the other hand, we can only rule out hot WD tertiaries at separations of $\gtrsim 2000$ au for Swift J1727, MAXI J1820, and GRO J0422. No meaningful constraints can be placed on WD tertiaries to V4641 Sgr and GS 2000. Furthermore, the cooling time for a WD to reach an effective temperature of $\lesssim 10^5$ K is only $\sim 10^6$ yr \citep[e.g.,][]{althaus_wd_2010}. Practically speaking, if any of these objects had a cool WD tertiary, we most likely would not have detected it.

\section{Discussion} 
\label{sec:discussion}

\subsection{Constraints on Gaia BH1 as a Triple}

Since Gaia BH1 is the best-studied object in our sample, we can place the most stringent constraints on its potential tertiaries. In \citet{el-badry_sun-like_2023}, spectral disentangling was used to rule out $\gtrsim 0.5\,M_{\odot}$ MS tertiaries to Gaia BH1. In \citet{nagarajan_espresso_2024}, high-precision radial velocity follow-up was used to place stringent limits on the hypothetical presence of an inner BH binary in Gaia BH1. Now, we can combine those results with the $5\sigma$ contrast curve for Gaia BH1 from this work to further constrain the allowed parameter space for MS tertiaries. 

\begin{figure*}
    \centering
    \includegraphics[width=\textwidth]{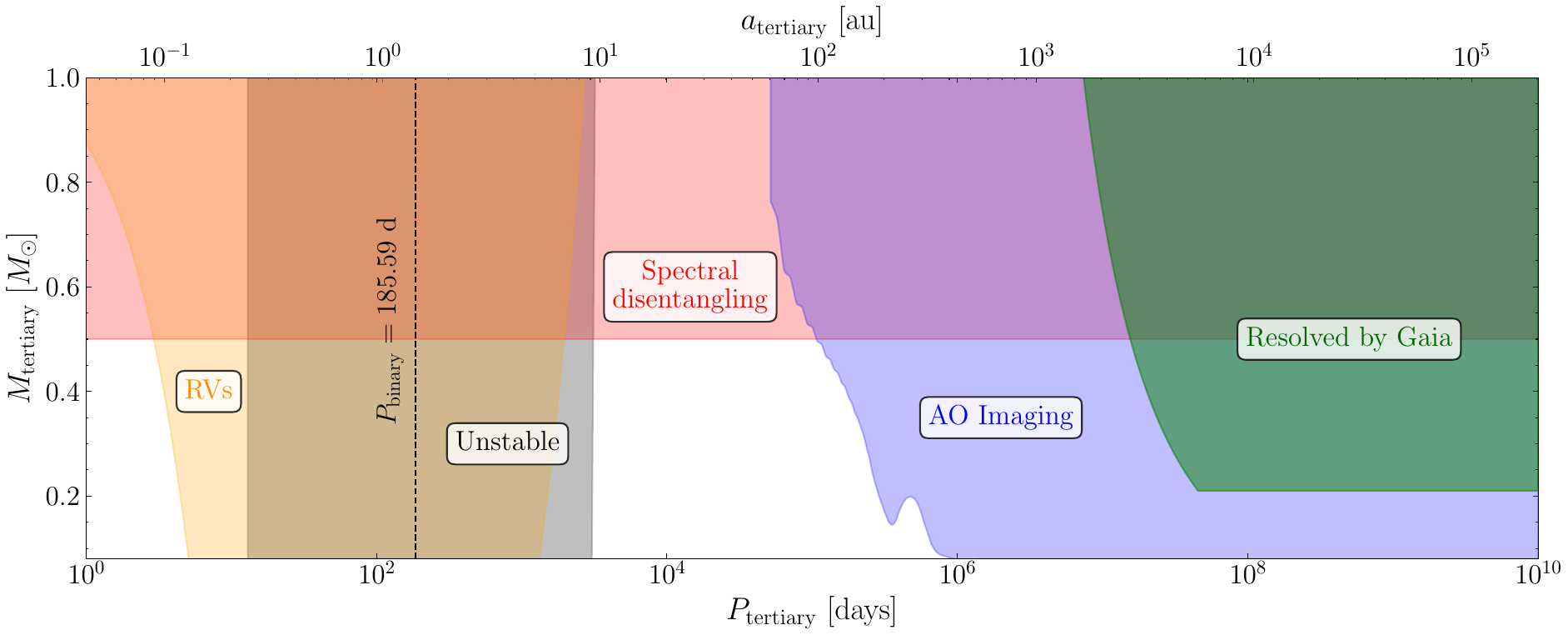}
    \caption{Constraints on main sequence tertiaries to Gaia BH1 from NIRC2-LGS AO imaging (blue, this work), precision radial velocities (RVs, orange) \citep{nagarajan_espresso_2024}, and spectral disentangling (red) \citep[][]{el-badry_sun-like_2023}. We also show regions of parameter space that are excluded by dynamical constraints (gray) \citep{mardling_aarseth_2001} and \textit{Gaia} observations (green). Given these constraints, only M dwarf (outer) tertiaries with mass $\lesssim 0.5\,M_{\odot}$ and orbital periods of $\sim 10^4$--$10^6$ d remain feasible. M dwarf (inner) tertiaries with orbital period $\lesssim 5$ d are also not ruled out by the RVs.}
    \label{fig:bh1_constraints}
\end{figure*}

We show the available constraints on MS tertiaries as a function of tertiary orbital period in Figure~\ref{fig:bh1_constraints}. We plot the combined constraints from spectral disentangling \citep{el-badry_sun-like_2023}, precision radial velocities (RVs) \citep{nagarajan_espresso_2024}, and NIRC2-LGS AO imaging (this work). We show the dynamically unstable region of parameter space, according to the hierarchical triple stability criterion of \citet{mardling_aarseth_2001}, as well. In doing so, we assume typical values for the outer eccentricity and mutual inclination of $0.5$ and $90^{\circ}$, respectively (i.e., based on the catalog of $\sim 10,000$ resolved triples constructed from \textit{Gaia} DR3 by \citealt{shariat_resolved_2025}). We also show the region of parameter space corresponding to tertiaries with apparent $G$-band magnitudes $> 20.7$ and angular separations $>1\arcsec$, and hence resolvable by \textit{Gaia}. We find that only M dwarf (outer) tertiaries with mass $\lesssim 0.5\,M_{\odot}$ and orbital periods of $\sim 10^4$--$10^6$~d (corresponding to outer separations of $\sim 10$--$400$ AU) would have evaded observational detection. M dwarf (inner) tertiaries with orbital period $\lesssim 5$~d, which are not ruled out by the RVs, could also exist. Cool WD (outer) tertiaries remain viable over a broader range of separations, as they are only constrained by the lower limit on orbital period set by dynamical stability.

\subsection{Implications for Formation Pathways}

\subsubsection{Gaia BHs and NSs}

We identify only 1 of the 12 observed Gaia NSs to be a hierarchical triple candidate, implying a tertiary fraction of $\lesssim 10\%$ for this population. While MS tertiaries have been proposed to aid in the formation of astrometric compact object binaries \citep[e.g.,][]{generozov_perets_2024, li_formation_2026}, our observations rule out most MS tertiaries at separations $\gtrsim 500$ au. On the other hand, the observed Gaia binaries could still have faint, close-in MS companions, or cool WD tertiaries that used to be intermediate-mass MS companions. These WD tertiaries would only be detectable at significantly higher contrasts, motivating deeper searches with follow-up imaging.

Gaia NSs are expected to receive small natal kicks, explaining their moderate eccentricities and wide orbits \citep[e.g.,][]{el_badry_2024}. That being said, even $\approx 10$--$20$ km s$^{-1}$ kicks are sufficient to unbind wide, loosely bound tertiaries \citep[e.g.,][]{burdge_nature_2024, shariat_cygni_2025}. Hence, these systems could have formed via hierarchical triple evolution in the past, but appear as isolated binaries today. Unfortunately, this alternative scenario is virtually untestable.

A different triple hypothesis is that Gaia BHs and NSs formed with a low-mass companion orbiting two massive stars. Since the massive stars would prevent one another from expanding to red supergiant dimensions, this could lead to systems with inner compact object binaries, potentially detectable with precision radial velocity follow-up \citep{hayashi_constraining_2023, nagarajan_espresso_2024}. Finally, the unseen compact object could have arisen from a merger product of the inner binary \citep{generozov_perets_2024, li_2024, naoz_links_2025}, which would produce a more top-heavy BH or NS mass function than expected from isolated binary evolution. Our observations do not constrain these alternative formation scenarios.

\subsubsection{BH LMXBs}

We do not identify any promising tertiary candidates for the quiescent, dynamically confirmed BH LMXBs in our sample, leaving V404 Cyg as the only confirmed hierarchical triple in this population to date. On the other hand, the fact that the companion to V404 Cyg is a relatively bright, evolved $\approx 1.2\,M_{\odot}$ star implies that this search is incomplete \citep[e.g.,][]{burdge_nature_2024}. A wider search of quiescent BH LMXBs (such as 1A 0620-00 or MAXI J1305-704), and deeper follow-up with vortex coronagraphy \citep[e.g.,][]{xuan_nirc2_2018, 2019AJ....157..118R, 2024AJ....168..215S} or interferometry, could build upon our initial survey. Measurement of proper motions over an extended time baseline (e.g., leveraging archival imaging) is also necessary to establish physical associations and constrain the overall tertiary fraction of this population.

The most likely tertiary separation, at least for BH LMXBs, is predicted to be $\sim 5000$ au \citep{naoz_2016, shariat_cygni_2025}. At that separation, we can rule out most MS companions. However, the most common companions are predicted to be WDs \citep{naoz_2016, shariat_cygni_2025}, which we would not have been able to detect.

Based on their kinematics, some BHs in binaries show evidence of a substantial natal kick, while others have properties disfavoring a large natal kick \citep[e.g.,][]{fragos_understanding_2009, atri_potential_2019, zhao_evidence_2023, vigna-gomez_constraints_2024, nagarajan_mixed_2024}. In our observed sample, Swift J1727.8-162 and V4641 Sgr show evidence of substantial natal kicks, while a significant natal kick is disfavored for MAXI J1820+070 \citep[e.g.,][]{nagarajan_mixed_2024}.\footnote{Robust proper motions are not yet available for GRO J0422+32 and GS 2000+25, precluding a natal kick analysis.} Thus, future follow-up should focus on BH LMXBs such as 1A 0620-00, which (like V404 Cygni) are more likely to have formed with weak kicks and hence have retained loosely bound tertiaries \citep[e.g.,][]{burdge_nature_2024, nagarajan_realistic_2025}.

\subsection{Future Prospects}

In order to prioritize the breadth of our survey (in light of the technical difficulties faced over the course of the campaign), we were unable to spend significant observing time leveraging other high-contrast imaging techniques such as angular differential imaging (ADI). This limits our achievable contrast at close separations. We reserved ADI follow-up for cases where we visually recognized a promising close tertiary candidate, since high contrast requires a significant amount of rotation (i.e., multiple hours of observation) near zenith \citep[e.g.,][]{xuan_nirc2_2018, 2019AJ....157..118R, 2024AJ....168..215S}. We applied this technique to Gaia NS1, finding that a bright speckle that we initially considered to be a promising tertiary candidate was actually an imaging artifact (see Appendix \ref{sec:adi}). 

Improved limits could be set using deeper observations with JWST; for instance, a 27.0 AB magnitude (23.8 Vega magnitude, \citealt{2016jdox.rept......}) WD can be detected at 10$\sigma$ significance in the NIRCam F444W filter in a 1 ks exposure \citep{2016jdox.rept......}. Adopting this detection limit, assuming negligible extinction, and using the same approach as in Section~\ref{sec:companion_constraints}, we find that JWST would be sensitive to any plausible WD companion (i.e, $T_{\text{eff}} > 3000$ K) out to a distance of $\approx 0.6$ kpc, and to WD companions of $T_{\text{eff}} \gtrsim 10^4$ K out to a distance of $\approx 1.8$ kpc. Very close-in companions can also be revealed with interferometry. VLT/GRAVITY+, which is designed to be assisted by AO, achieves a spatial resolution of at least 4 mas in the $K$-band \citep{gravity_first_2017}. Finally, additional imaging from Roman's Galactic Plane Survey (GPS) could enable a common proper motion analysis to determine if tertiary candidates are gravitationally bound. The GPS will cover most of the Galactic plane, reaching a detection limit of $21.18$ Vega magnitude in the WFI F213 filter \citep{2022rdox.rept......}. Furthermore, the GPS is expected to deliver proper motion measurements at a precision similar to that of \textit{Gaia} at \textit{Gaia}'s detection limit, but will also extend to fainter magnitudes at wavelengths less affected by extinction \citep{2022rdox.rept......}.

\section{Conclusion}
\label{sec:conclusion}

We used Keck/NIRC2 with laser guide star adaptive optics (AO) to acquire deep near-infrared images of five quiescent BH low-mass X-ray binaries (LMXBs), Gaia BH1, and twelve Gaia NSs. By searching for candidate tertiaries, we test hierarchical triple formation scenarios for these systems, which isolated binary evolution models struggle to produce. We summarize the main results of our campaign below.

\begin{itemize}
    \item We detect numerous faint sources that were previously unresolved in survey imaging (Figures~\ref{fig:comparison}, \ref{fig:narrow_images}, and \ref{fig:wide_images}), but cannot robustly confirm any of them as bound companions. We identify a few promising tertiary candidates with chance alignment probabilities $\lesssim 5\%$ (Table~\ref{tab:ao_cands}), but none of the candidates identified in the \textit{Gaia} DR3 catalog have proper motions or parallaxes consistent with their associated targets. One source (estimated to be an M4V star) located $\approx 4.3\arcsec$ to the west of J1733+5805 (a Gaia NS) is undetected in DR3 and remains a potential tertiary candidate, but further follow-up is necessary to measure proper motions and establish a physical association.  
    
    \item To search for tertiaries at close separations ($\lesssim 1\arcsec$), we used the reference star differential imaging (RDI) strategy together with the Karhunen-Lo\'eve Image Processing algorithm to model and subtract the PSFs of our targets and achieve high contrast sensitivity at close separations (typically, $\Delta K_s = 5$ mag at an angular separation of $0.5\arcsec$) (Figures~\ref{fig:narrow_rdi} and \ref{fig:wide_rdi}). While we detected several 5$\sigma$ sources (Table~\ref{tab:pca_cands}), visual inspection and injection-recovery tests suggest that they are consistent with being artifacts arising from residual speckle noise or radial diffraction spikes.
    
    \item To place constraints on undetected tertiaries, we computed $5\sigma$ contrast curves for each of our targets, representing the detection limit as a function of angular separation in our high contrast images (Figures~\ref{fig:narrow_contrast} and \ref{fig:wide_contrast}). From these contrast curves, we inferred maximum masses for main sequence (MS) tertiaries (Figure~\ref{fig:ms_constraints}) and maximum effective temperatures for white dwarf (WD) tertiaries (Figure~\ref{fig:wd_constraints}) that would be consistent with a non-detection. In general, we rule out plausible (i.e., above the hydrogen burning limit) MS tertiaries and very hot (i.e., $\gtrsim 10^5$ K) WD tertiaries to the Gaia compact object binaries at projected separations $\gtrsim 500$ au, and to the more distant BH LMXBs at projected separations $\gtrsim 2000$ au.
    
    \item The discovery that V404 Cygni is in a hierarchical triple with a relatively bright, evolved $\approx 1.2\,M_{\odot}$ companion lends credence to the hypothesis that hierarchical triple evolution plays an important role in the formation of BH LMXBs \citep[e.g.,][]{burdge_nature_2024, shariat_cygni_2025}. However, our constraints suggest that hierarchical triple models that predict present-day MS tertiaries do not represent the primary formation channel for these systems. Our observations leave open the possibility of undetected cool WD tertiaries, with deeper observations (e.g., with JWST) needed to reveal their presence. The dominant formation channel for Gaia BHs and NSs remains unclear as well, with a combination of constraints from AO imaging, precision radial velocities, spectral disentangling, and dynamical stability ruling out a large region of parameter space for MS tertiaries to Gaia BH1 (Figure~\ref{fig:bh1_constraints}).
\end{itemize}

In the future, common proper motion tests conducted over a time baseline of several years (i.e., based on follow-up imaging at later epochs) could confirm or rule out the tertiary candidates presented in this work. In addition, angular differential imaging, vortex coronagraphy, or interferometry could be used to improve constraints on close-in companions. 

%% Please use the acknowledgment and contribution environments. This will 
%% be anonomyized when the "anonymous" style option is used. 
\begin{acknowledgments}
We thank Natasha Abrams and Jerry Xuan for useful discussion. This research was supported by NSF grants AST-2307232 and AST-2540180. This research was supported in part by grant NSF PHY-2309135 to the Kavli Institute for Theoretical Physics (KITP).  A.S. acknowledges support from the National Science Foundation Graduate Research Fellowship under Grant No.~2139433. This work has made use of data from the European Space Agency (ESA) mission {\it Gaia} (\url{https://www.cosmos.esa.int/gaia}), processed by the {\it Gaia}
Data Processing and Analysis Consortium (DPAC,
\url{https://www.cosmos.esa.int/web/gaia/dpac/consortium}). Funding for the DPAC has been provided by national institutions, in particular the institutions
participating in the {\it Gaia} Multilateral Agreement. 
\end{acknowledgments}

\begin{contribution}
%%This section gives authors the space to recognize author contributions. The text inside this environment is NOT counted towards the total word quanta. At a minimum, manuscripts are expected to include this text:

PN was responsible for leading the observational campaign, performing the data analysis, and writing the manuscript. KE came up with the initial research concept and obtained the funding. AS provided insight on the use of the \texttt{VIP} package to perform reference star differential imaging and construct contrast curves.

%% But authors are expected to provide more specific details, e.g. 
%%
%%SC was responsible for writing and submitting the manuscript.
%%WWM came up with the initial research concept and edited the manuscript.
%%OTS obtained the funding and edited the manuscript.
%%EBF provided the formal analysis and validation. He also edited the manuscript.
%%GEH Supervised the undergraduates, wrote the software and administers the project github and Zenodo repositories.
%%
%% Authors can use the Contributor Role Taxonomy (CRediT) at
%% for ideas on how write a good statement tailored to their needs.

\end{contribution}

\facilities{Keck:II (NIRC2-LGS)}

\software{astropy \citep{2013A&A...558A..33A, 2018AJ....156..123A, 2022ApJ...935..167A}, \texttt{pyKLIP} \citep{pyklip_2015}, \texttt{VIP} \citep{gomez_vip_2017, 2023JOSS....8.4774C}}


\clearpage

\appendix

\section{Angular Differential Imaging of Gaia NS1}
\label{sec:adi}

\begin{figure}
    \centering
    \includegraphics[width=\columnwidth]{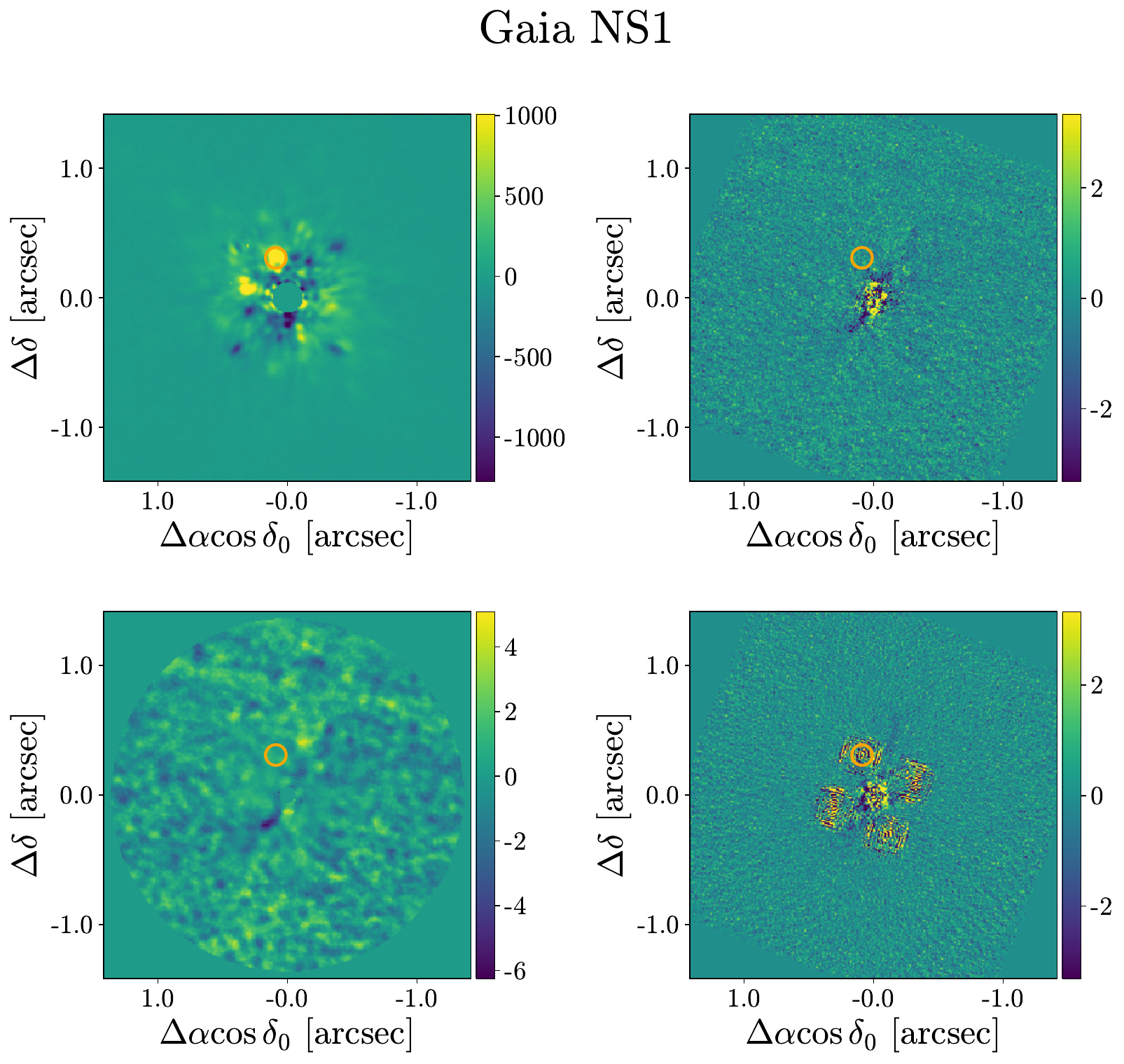}
    \caption{Investigation of a candidate tertiary identified at SNR $> 5$ in our NIRC2-LGS imaging of Gaia NS1 (upper left panel), marked with an orange circle. We pursued follow-up angular differential imaging and performed post-processing with the KLIP algorithm \citep{soummer_klip_2012} with 16 principal components (upper right panel). We do not detect any SNR $ > 5$ sources in the signal-to-noise ratio (SNR) map of the PSF-subtracted image (lower left panel), implying that the candidate was actually a quasi-static speckle. We perform PSF injection-recovery tests (lower right panel), finding that an astrophysically significant source as bright as the candidate would have been visually prominent, despite significant self-subtraction.}
    \label{fig:ns1_adi}
\end{figure}

We performed angular differential imaging (ADI) to investigate a close tertiary candidate to the northeast of Gaia NS1, visually identified during our observing campaign, marked with an orange circle in the upper left panel of Figure~\ref{fig:ns1_adi}. We obtained 48 exposures of Gaia NS1 on January 28, 2025 (UTC) with the narrow camera, using 6s exposures and a 5-point dither pattern. The total position angle rotation achieved was $\approx 22^{\circ}$. Principal component analysis (PCA) with the KLIP algorithm \citep{soummer_klip_2012} reveals that the candidate, located $\approx 0.3\arcsec$ northeast from the target, was actually a quasi-static speckle (upper right panel of Figure~\ref{fig:ns1_adi}). Specifically, we do not detect any SNR $> 5$ sources in our post-processed image obtained using ADI (lower left panel of Figure~\ref{fig:ns1_adi}). PSF injection-recovery tests at various position angles (lower right panel of Figure~\ref{fig:ns1_adi}) confirm that an astrophysically real source as bright as the identified speckle would have been visually prominent, even after significant self-subtraction.

\section{Injection-Recovery Tests}
\label{sec:injection_recovery}

\begin{figure*}
    \centering
    \includegraphics[width=\textwidth]{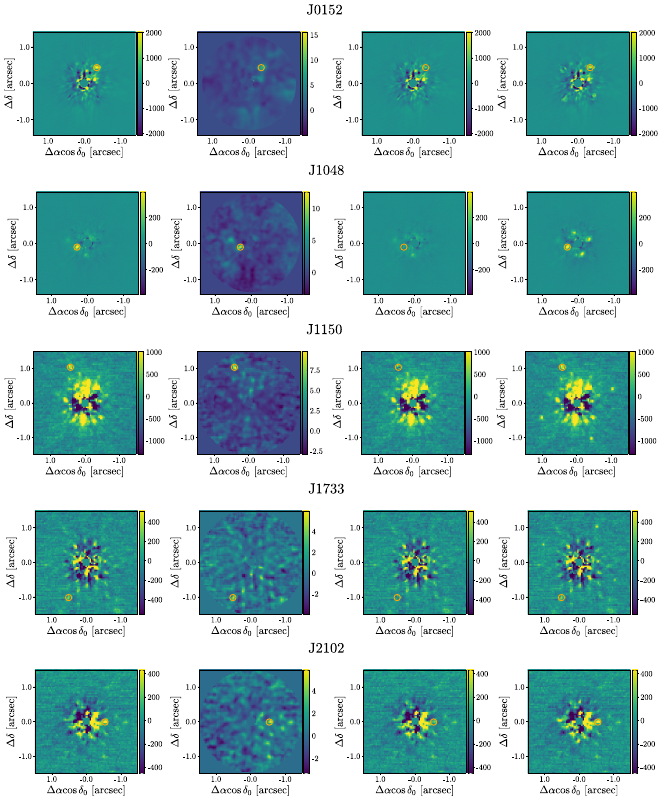}
    \caption{Injection-and-recovery tests for close tertiary candidates identified via RDI (Table~\ref{tab:pca_cands}). The first column shows the cropped science images, the second column shows the SNR maps, the third column shows the post-processed images with the candidates removed, and the fourth column shows the post-processed images with examples of injected fake companions. In each row, the location of the tertiary candidate is marked with an orange circle.}
    \label{fig:injection_recovery}
\end{figure*}

To check whether the tertiary candidates identified in Table~\ref{tab:pca_cands} are artifacts, we performed PSF injection-recovery tests. Specifically, we used the negative fake companion \citep[NEGFC;][]{2010Sci...329...57L, 2010Natur.468.1080M, wertz_negfc_2017} technique with the Nelder-Mead simplex algorithm to robustly estimate the flux and location of each candidate. Then, we removed this candidate from the science image. We injected fake companions of the same brightness into the science image, placing them at various position angles (PAs) at the candidate's measured separation. Finally, for each fake companion, we re-ran our reference star differential imaging (RDI) analysis and checked whether the source is recovered at 5$\sigma$ significance.

The results of this procedure are shown in Figure~\ref{fig:injection_recovery}. In each row, the first panel shows the cropped science image, the second panel shows the signal-to-noise ratio (SNR) map of the PSF-subtracted image, the third panel shows the PSF-subtracted image with the candidate removed, and the fourth panel shows the PSF-subtracted image with examples of injected fake companions (i.e., at 90$^{\circ}$ intervals in PA). In all panels, the candidate's location is marked with an orange circle. The PSF-subtracted images are stretched to emphasize the detected candidates, resulting in saturation of the central speckle regions.

Only the candidates for J1150 and J1733 are recovered at all or most injected PAs. However, from visual inspection, we find that these candidates are aligned with radial diffraction spikes. The candidates for J0152, J1048, and J2102 are recovered at $< 50\%$ of the injected PAs, suggesting that they arise from residuals due to imperfect PSF subtraction. Indeed, from visual inspection, we find that these candidates appear similar to quasi-static speckles. This exercise suggests that all close tertiary candidates identified from our RDI analysis are likely artifacts, though it is not possible to definitively rule out an astrophysical origin at this stage. Future high-contrast imaging follow-up observations can clarify the true nature of these sources by re-detecting them and confirming them as bound companions using a common proper motion test.

\clearpage

%% For this sample we use BibTeX plus aasjournalv7.bst to generate the

\bibliography{bibliography}{}
\bibliographystyle{aasjournalv7_modified}



\end{document}